\documentclass[fleqn,usenatbib]{mnras}

\usepackage{newtxtext,newtxmath}
\usepackage[T1]{fontenc}

\DeclareRobustCommand{\VAN}[3]{#2}
\let\VANthebibliography\thebibliography
\def\thebibliography{\DeclareRobustCommand{\VAN}[3]{##3}\VANthebibliography}

\usepackage{graphicx}	% Including figure files
\usepackage{amsmath}	% Advanced maths commands
\usepackage{bm}
\usepackage{hyperref}
\usepackage{graphicx}
\usepackage{xspace}
\usepackage{url}
\title[True repeating FRB fraction]{The true fraction of repeating fast radio bursts revealed through CHIME source count evolution
}

\author[Yamasaki et al.]
{Shotaro Yamasaki\thanks{E-mail: shotaro.s.yamasaki@gmail.com}$^{1}$, Tomotsugu Goto$^{2,3}$, Chih-Teng Ling$^{3}$ and Tetsuya Hashimoto$^{1}$
\\
$^{1}$Department of Physics, National Chung Hsing University, 145 Xingda Rd., South Dist., Taichung 40227, Taiwan (R.O.C.)\\
$^{2}$Institute of Astronomy, National Tsing Hua University, 101, Section 2. Kuang-Fu Road, 
Hsinchu, 30013, Taiwan (R.O.C.)\\
$^{3}$Department of Physics, National Tsing Hua University, 101, Section 2. Kuang-Fu Road, 
Hsinchu, 30013, Taiwan (R.O.C.)
}

\date{Accepted XXX. Received YYY; in original form ZZZ}

\pubyear{2023}

\begin{document}
\label{firstpage}
\pagerange{\pageref{firstpage}--\pageref{lastpage}}
\maketitle

\begin{abstract}
%%% Short Version %%%
Fast Radio Bursts (FRBs) are classified into repeaters and non-repeaters, with only a few percent of the observed FRB population from the Canadian Hydrogen Intensity Mapping Experiment (CHIME) confirmed as repeaters.  However,  this figure represents only a lower limit due to the observational biases, and the true fraction of repeaters remains unknown.  Correcting for these biases uncovers a notable decline in apparently non-repeating FRB detection rate as the CHIME operational time increases. This finding suggests that a significant portion of apparently non-repeating FRBs could in fact exhibit repetition when observed over more extended periods. A simple population model infers that the true repeater fraction likely exceeds 50\% with 99\% confidence, a figure substantially larger than the observed face value, even consistent with 100\%. This greater prevalence of repeaters had previously gone unnoticed due to their very low repetition rates ($\sim$10$^{-3.5}$ hr$^{-1}$ on average). 
Hence, theoretical FRB models must incorporate these low-rate repeaters. Furthermore, our results indicate a significantly higher repeater volume number density, potentially exceeding observed values by up to 10$^4$ times, which in turn impacts comparisons with potential FRB progenitors.

\end{abstract}

% Don't make up new ones.s

%%%%%%%%%%%%%%%%%%%%%%%%%%%%%%%%%%%%%%%%%%%%%%%%%%

%%%%%%%%%%%%%%%%% BODY OF PAPER %%%%%%%%%%%%%%%%%%

\section{Introduction} \label{sec:intro}
Fast Radio Bursts (FRBs) are mysterious transient events characterized by intense and brief flashes of radio emissions \citep{lorimer07,petroff22}. Observationally, FRB sources are classified into two categories: (apparently) non-repeating or repeating sources. This classification is inevitably incomplete because, within the limited observing period, we do not know whether the apparent non-repeater will repeat or not in the future \citep{caleb19,james20a}. In this sense, the observed fraction of repeaters (approximately a few percent; \citealt{chime-repeaters23}) is merely a lower limit. 

Moreover, observations are also incomplete. 
For example, 
the exposure time of the Canadian Hydrogen Mapping Experiment (CHIME) heavily depends on the declination, owing to its geographic location and configuration. This results in shorter observation times for declination angles near the telescope's southern limit compared to those further north \citep{chime-catalog1}. Consequently, the amount of time spent observing a given sky position varies significantly over declination. 
While this facilitates separating the sky into several bins and derives the burst detection rate only within those bins with similar exposure times \citep{chime-repeaters23}, it does limit the potential for statistical analysis and uniform treatment of all bursts.

Here, we overcome this limitation, by correctly accounting for the declination dependence of exposures and time gaps between daily exposures. After applying these corrections, one can use FRB population models \citep[e.g.,][]{gardenier19,lu20,ai21} with the observed, corrected source count evolution to infer FRB population characteristics and hidden repetition properties. This enables us to accurately determine the true fraction of repeaters. 

The paper is structured as follows: In \S \ref{sec:observed sc}, we outline the correction process for the observed CHIME source counts. In \S \ref{sec:model}, we detail the source count modeling. Finally, the implications of our findings are discussed in \S \ref{sec:discussion}.

\section{Correcting Observed Source Counts} 
\label{sec:observed sc}
We use the FRB data from CHIME, a radio interferometer, which performs daily observations of the entire Northern sky ($\delta>-11^\circ$). The CHIME/FRB Project has released its first catalog of bursts \citep{chime-catalog1}, comprising the largest sample of distinct FRB sources detected by a single instrument. This extensive catalog provides an excellent opportunity for a comprehensive investigation into the evolution of FRB source counts over time.
For our study, we utilize 393 bursts detected during the second half of the operational period (between December 31, 2018, and  July 1, 2019)\footnote{We have excluded 25 newly identified repeating FRB sources \citep{chime-repeaters23}, which were detected using a different algorithm during extended observation periods, to mitigate potential impacts stemming from non-uniform sensitivity. }, when the telescope sensitivity was more stable after the initial launching period (Appendix \ref{app:period selection}). 

To correct for the aforementioned observational biases, 
we introduce the quantity $f_{\rm exp}=T_{\rm exp}/T$, which represents the fraction of the total exposure time ($T_{\rm exp}$) spent on a specific burst since December 31, 2018, at 00:00:00 universal time ($T_0$), until its detection and $T$ is the total CHIME operation time during the same period. Specifically, $f_{\rm exp}$ is 1 at $\delta=90^\circ$. We then assign a weight given by $1/f_{\rm exp}$ to each burst. 
By summing these weights, we obtain the correct source counts as a function of $T$ for both repeaters and apparently non-repeating sources, effectively correcting the declination-dependent exposure times (Appendix \ref{app:observed sc}).

\begin{figure}
\centering
\includegraphics[width=0.48\textwidth]{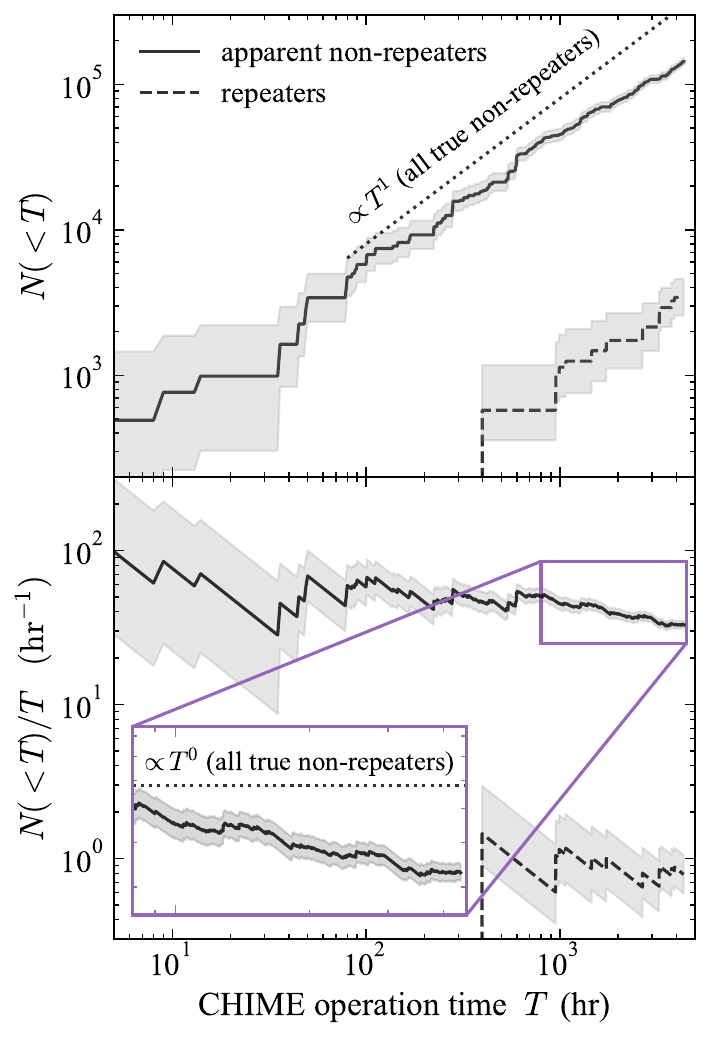}
\caption{The observed cumulative FRB source counts as a function of CHIME operation time. {\it Top panel}: the exposure-corrected cumulative source counts. The solid and dashed lines represent apparently non-repeating and repeating source counts, respectively.  {\it Bottom panel}: average detection rates (cumulative source counts per operational time in units of hours). The inset illustrates a zoomed-in view of a decreasing trend in the apparently non-repeating source detection rate at late operation times, specifically at $\gtrsim10^3$ hr. In both panels, the CHIME operation time has been measured since $T_0$, and the dotted black line demonstrates the expected evolution of the apparently non-repeating source count when the entire population consists of true non-repeating sources.  
}
\label{fig:sc_obs}
\end{figure}

Fig.~\ref{fig:sc_obs} 
displays the source counts for both observed apparently non-repeating (the solid line) and repeating (the dashed line) populations as a function of CHIME operation time since $T_0$. 
All FRBs are regarded as non-repeaters until their second burst is detected when they are re-classified as repeaters.
The observed apparently non-repeating source detection rate (bottom panel of Fig~\ref{fig:sc_obs}) shows an almost constant value during early operation times, followed by a significant decrease after $T\gtrsim 500$ hr by about a factor of two until the end of the observation period. 
The decrease in the apparently non-repeating source count rate is due to the detection of the second burst from repeating sources, resulting in the re-classification of previously non-repeating sources into repeating sources. As a result, the count rate of apparently non-repeating sources effectively decreases over time, as more and more repeaters are correctly classified as such.
This continued decrease of the apparently non-repeating source counts after long operation times of $>$1000 hours indicates the existence of still misclassified repeaters due to their low repeat rates.

In contrast, the repeating source count in the dashed line begins to appear at $T\sim 500$ hr, indicating that on average, such operation time is required to begin detecting the noticeable number of second bursts from repeaters.
The repeater's detection rate remains constant until the end of the operation period, implying that repeaters have a broad repeating rate distribution.  Those low-rate repeaters continue to exist up to $T\sim4000$ hr, or possibly beyond.

\section{Modeling Source Count Evolution} 
\label{sec:model}
To qualitatively interpret the results, we simulate FRB populations based on a simple model, to reproduce detection and classifications under the CHIME observing setups as a function of $T$ (Appendix \ref{app:simulation}).
Because our primary goal is to interpret the observed source counts, we adopted a simplified approach, focusing on modeling the observed FRB population above CHIME's detection limit ($\sim 5$ Jy ms; \citealt{chime-catalog1}). 
We employ a Monte Carlo simulation that includes both repeating and truly non-repeating sources\footnote{Throughout the paper, we focus on the observational properties of FRBs, and the use of the term ``true'' (or ``genuine'') repeating/non-repeating FRBs is not intended as a physical definition but rather as a descriptive label within the observational context.}. In the case of repeating sources, we model the distribution of time intervals between consecutive bursts using the Weibull distribution, based on the evidence for highly clustered distribution in the burst arrival times inferred from some repeaters \citep{opperman18,nimmo23}. The waiting time distribution is characterized by two parameters: the mean repeating rate denoted as $r$, and the degree of wait-time clustering represented by $k$. 
To determine the mean repeating rate ($r$), we sample from the differential rate distribution $dN/dr \propto r^{-q}$ $(r_{\rm min}<r<r_{\rm max})$, where $q$ ($>0$) is the power-law index and $r_{\rm min}$ is the minimum repeating rate. Both $q$ and $r_{\rm min}$ are free parameters in our analysis. We set the maximum repeating rate $r_{\rm max}$ to a fixed value of $10^{0.5}\ {\rm hr^{-1}}$ \citep{ai21,lu20}, constrained by the most active repeaters in the literature. 

Under the assumption of uniformly distributed repeating sources across the sky ($-11^\circ < \delta < 90^\circ$) and burst time series following the Weibull distribution, we simulate the hypothetical CHIME detections, considering the declination-dependent exposure time window for each transit. Within these simulated time windows, we count apparently non-repeating ($N_{\rm rn}$, only detected once so far) and repeating  ($N_{\rm rr}$, detected more than once) events from the repeating population. We introduce the total number of simulated repeating sources, denoted as $N_{\rm r}^{\rm tot}$, as a free parameter that scales with the model source count (Appendix \ref{app:simulation}).

\begin{figure}
\centering
\includegraphics[width=0.48\textwidth]{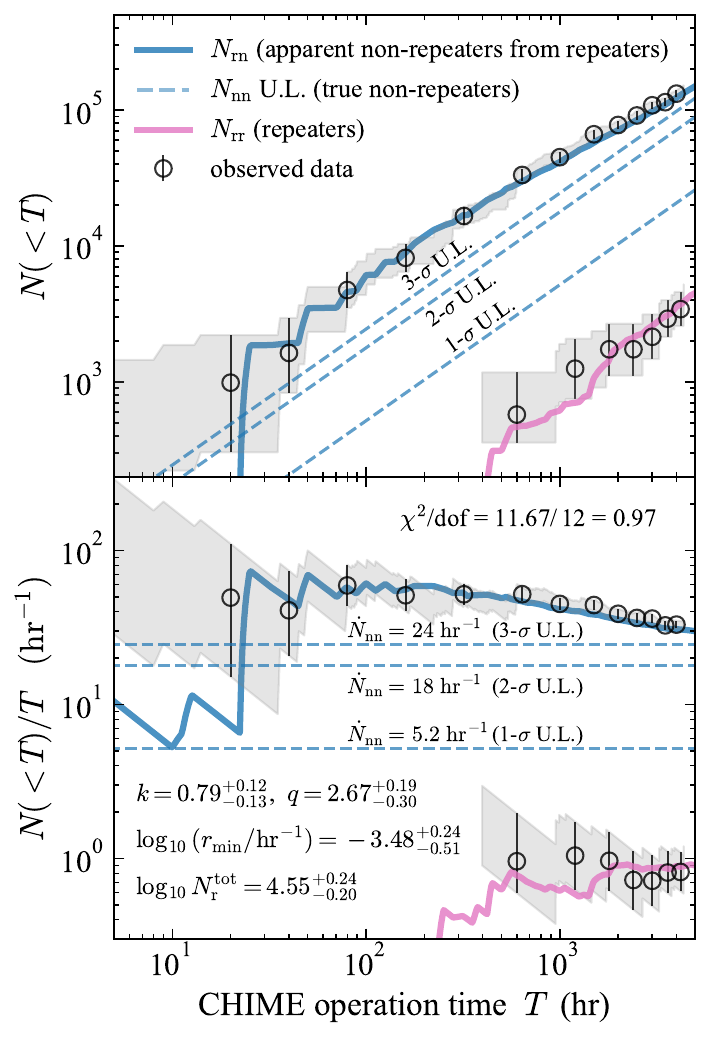}
\caption{Model fit to the observed cumulative FRB source counts.  The observed data in the black open circles are extracted from the black lines in Fig. \ref{fig:sc_obs}. The blue solid line shows the best-fit apparently non-repeating source counts. The pink solid curve corresponds to the best-fit repeating source count. The dashed blue lines represent the 1--3 $\sigma$ upper limits on the apparently non-repeating source count originating from truly non-repeating populations. The reduced $\chi^2$ value and best-fit parameters are indicated in the lower panel (see also Fig. \ref{fig:mcmc_corner}). }
\label{fig:sc_fit}
\end{figure}

The source count of genuine non-repeating sources is expected to increase linearly with telescope operation time $T$ at a constant event rate density \citep{ai21}, as shown in the dotted line in Fig.~\ref{fig:sc_obs}. We model the cumulative apparently non-repeating source count of truly non-repeating sources by $N_{\rm nn}=\dot{N}_{\rm nn} T$, where $\dot{N}_{\rm nn}$ is the cosmic non-repeating burst rate, a free parameter in our model. 

We fit the observed cumulative repeating source count by $N_{\rm rr}$ and the cumulative source count of apparently non-repeating sources by $N_{\rm rn}+N_{\rm nn}$ (Appendix \ref{app:mcmc}).
After the Markov Chain Monte
Carlo (MCMC) fit, the observed source counts are well-explained by the model shown in Fig.~\ref{fig:sc_fit} with best-fit parameters for repeating population: $\log_{10}(r_{\rm min}/{\rm hr^{-1}})=-3.48_{-0.47}^{+0.23}$, $q=2.67_{-0.29}^{+0.17}$, $k=0.79_{-0.12}^{+0.12}$, and $\log_{10} N_{\rm r}^{\rm tot}=4.55_{-0.19}^{+0.24}$
(at 1-$\sigma$ confidence level). 
The best-fit result shows a remarkably low upper limit on the genuine non-repeater rate: 
$\dot{N}_{\rm nn}< 5.2\ {\rm hr^{-1}}$ (1$\sigma$), implying that the majority of apparent non-repeaters will likely be re-classified as repeaters with sufficiently long exposure time (see blue dashed line in the lower panel of Fig. \ref{fig:sc_fit}). 
The corrected fraction of repeaters denoted as $f_{\rm r}=(N_{\rm rr}+N_{\rm rn})/N_{\rm tot}$, exceeds 50\%, where $N_{\rm tot}=N_{\rm rr}+N_{\rm rn}+N_{\rm nn}$ represents the total source counts. This value is significantly higher than the observationally confirmed repeating source fraction of $N_{\rm rr}/N_{\rm tot}$=$2.6^{+2.9}_{-2.6}$\% \citep{chime-repeaters23}. i.e., among 393 FRBs in the catalog, at least $\sim$ 200 may be actually repeaters, while only 15 are known repeaters.

Furthermore, the estimated $f_{\rm r}=50\%$ is most likely a lower limit. In Fig. 1, the ratio $N/T$ of non-repeaters continues to decline towards the end of the observation period at 4000 hrs. If $N/T$ continues to decline at $T>4000$ hrs, $f_{\rm r}$ would be even larger. Indeed, the MCMC likelihood (Fig. \ref{fig:mcmc_corner}) shows that $f_{\rm r}$ is consistent to be 100\% within one sigma. If this is the case, all FRBs could be repeaters after a sufficiently long exposure time. 

The inferred minimum repeating rate for repeaters stands at $r_{\rm min}\sim 10^{-3.5}$ hr$^{-1}$, a value notably consistent with CHIME's operational timescale of 4000 hours. In principle, the minimum repeating rate is expected to degenerate with the event rate for non-repeating FRBs. This is because distinguishing between non-repeating sources and repeating sources with mean time intervals exceeding the current observing time can be challenging. 
However, we ensure that our fitting result is not influenced by the observation timescale (see Appendix \ref{app:assessment}). Instead, it reflects the true repetition properties. Qualitatively, this minimum repeating rate is determined by the interplay between the appearance timescale of the diminishing trend in non-repeaters and the emergence of repeating sources, as well as the source count ratio between repeaters and non-repeaters. Analysis of the current CHIME data suggests an approximate timescale of $10^{3.5}$ hours for this interplay.

\section{Implications} \label{sec:discussion}

Through a comparison of the FRB event number density with potential progenitors, a previous study \citep{ravi19} claimed that FRBs cannot solely arise from a non-repeating population.
In this work, we present direct observational evidence demonstrating that the fraction of repeaters is consistently growing, accurately establishing that the true non-repeater fraction remains below 50\%, solely relying on FRB source count information.

\begin{figure}
\centering
\includegraphics[width=0.48\textwidth]{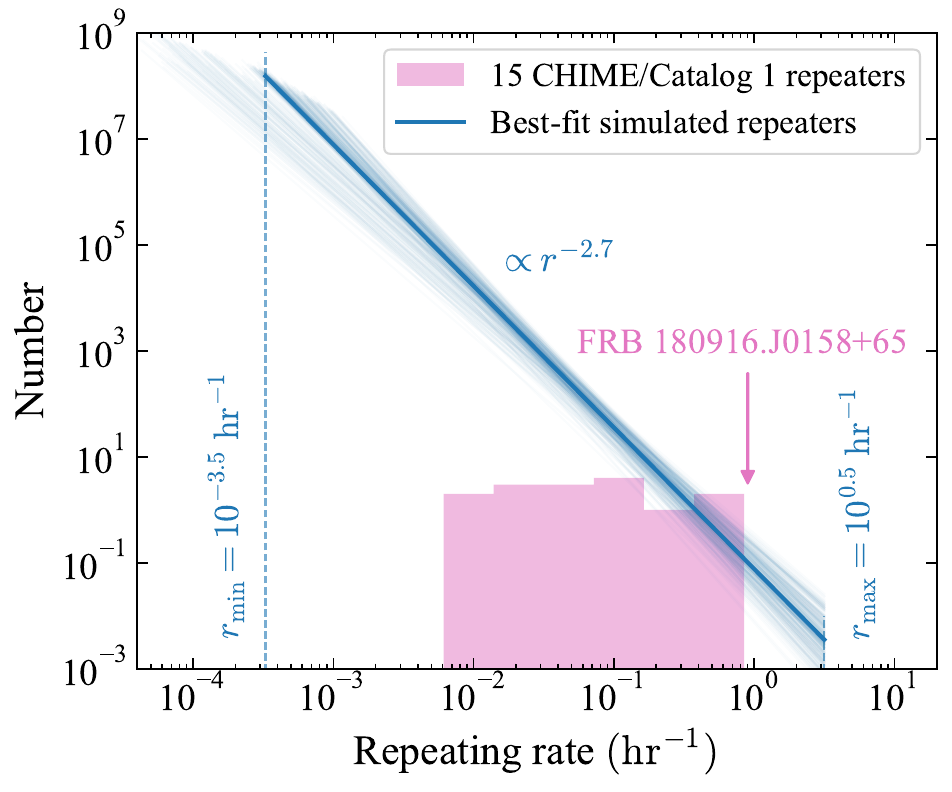}
\caption{Repeating rate distribution: model vs. CHIME repeater data.  The blue lines represent the best-fit repeating rate distribution obtained from our model, which takes into account non-Poissonian wait-time clustering and CHIME observing setups. The pink histogram shows the repeating rate distribution of the observed CHIME repeaters that were confirmed during the Catalog 1 period  \citep{chime-repeaters23}. The pink downward arrow points to the most active repeating source, FRB 180916.J0158+65  \citep{chime-repeaters19,chime-R2-20}}
\label{fig:rate_comparison}
\end{figure}

The best-fit repeat rate distribution is shown in Fig.~\ref{fig:rate_comparison}.
Note that the distribution based on the best-fit parameters (blue solid line) includes simulated FRB sources that are not yet detected by CHIME, not even once as an apparent non-repeater. This represents the true number of sources above the CHIME detection limit throughout the Universe, irrespective of human detection. In comparison to the known repeaters (pink histogram), the best-fit power-law model shows a much larger number of repeaters with a low repeating rate. This striking difference in repeating rates suggests that CHIME is preferentially detecting relatively rare and frequently repeating sources, such as an active CHIME repeater FRB 180916.J0158+65  \citep{chime-repeaters19,chime-R2-20}, 
while the majority of low-repetition rate ($\sim10^{-3.5}$ times per hours) sources are yet to be found. 

A substantially larger repeater fraction $f_{\rm r}$ suggested in this work would have a profound impact on studies comparing the number density of FRBs to that of other astronomical objects to constrain their progenitors \citep{palaniswamy18,ravi19,james20b,hashimoto20,luo20}.
Considering the prevalence of repeating FRBs with low repeating rates, we revise the local volume density of repeating FRB sources to be $n_{\rm FRB}=1.2^{+0.9}_{-0.4}\times 10^{7}$ Gpc$^{-3}$ (Appendix \ref{app:number density}). 
Our estimate is in agreement with the previous work using dispersion measure distribution giving a broad range of 10$^4$--10$^{10}$ Gpc$^{-3}$ \citep{james23}, but our approach provides a more accurate constraint due to its utilization of the source count evolution. 
This evolution captures the entire time evolution throughout the observation period, providing unique information for constraining the actual number density of repeating FRB sources.
The increased $n_{\rm FRB}$ suggests the need for larger $\tau_{\rm age}$ to be compatible with existing progenitor models. For instance, the birth rate of neutron stars from core-collapse supernovae is $10^5$ Gpc$^{-3}$ yr$^{-1}$, requiring $\tau_{\rm age}\gtrsim10^2$ yr.
In any case, the number density arguments must carefully account for the potentially larger presence of undetected repeaters, particularly those with low repeat rates. In this regard, ongoing and future all-sky FRB surveys \citep{stare2,grex,burstt} offer the most efficient means to address this subject.

Physical origins of repeaters and apparent non-repeaters have often been discussed separately \citep{zhang20}. 
Our findings, which point to the existence of a significantly overlooked population of low-rate repeaters, question this interpretation. The presence of more infrequent repeaters may cast doubt on the previously suggested dichotomy between repeaters and apparent non-repeaters \citep{chime-repeaters19,fonseca20,chime-catalog1}, thus challenging our understanding of the diversity within the FRB population.

Lastly, we acknowledge the independent work conducted by \citet{mcgregor23}, who performed a similar analysis on the evolution of CHIME repeater source counts. The primary difference is that their focus is solely on repeaters. Remarkably, their conclusion that approximately 170 currently non-repeating sources in CHIME Catalog 1 will repeat well aligns with our prediction of about 200.

\section*{Acknowledgements}
This work made use of publicly available data from CHIME, the Canadian Hydrogen Mapping Experiment. SY thanks Hsiu-Hsien Lin for the valuable assistance in providing the CHIME/FRB Catalog 1 data details. We thank Ziggy Pleunis and Kiyo Masui for making the information on the CHIME exposure time history as a function of CHIME operation time publicly available at \url{https://github.com/chime-frb-open-data/community/issues/95}. We also thank the referee for
providing a valuable suggestion to assess the validity of our fitting result, which greatly helped improve the quality of the manuscript.
TG acknowledges support from the National Science and Technology Council of Taiwan through grants 112-2112-M-007 -013, and 112-2123-M-001 -004 -. 
TH acknowledges support from the National Science and Technology Council of Taiwan through grants 110-2112-M-005-013-MY3, 110-2112-M-007-034-, and 112-2123-M-001-004-.

%%%%%%%%%%%%%%%%%%%%%%%%%%%%%%%%%%%%%%%%%%%%%%%%%%

\section*{Data Availability}
All data analyzed in the current study are available from the CHIME/FRB Public Database (\url{https://www.chime-frb.ca/}) and CHIME/FRB Open Data (\url{https://chime-frb-open-data.github.io/}). Custom codes and the files to
reproduce source count analysis will be made available upon reasonable request to the corresponding author. %\url{https://github.com/shotaro-yamasaki/frbsc}.



\bibliographystyle{mnras} 
\input{main.bbl}
% if your bibtex file is called example.bib

%%%%%%%%%%%%%%%%%%%%%%%%%%%%%%%%%%%%%%%%%%%%%%%%%%

%%%%%%%%%%%%%%%%% APPENDICES %%%%%%%%%%%%%%%%%%%%%

\appendix

\section{Selection of CHIME operation period}
\label{app:period selection}

\begin{figure}
\centering
\includegraphics[width=0.48\textwidth]{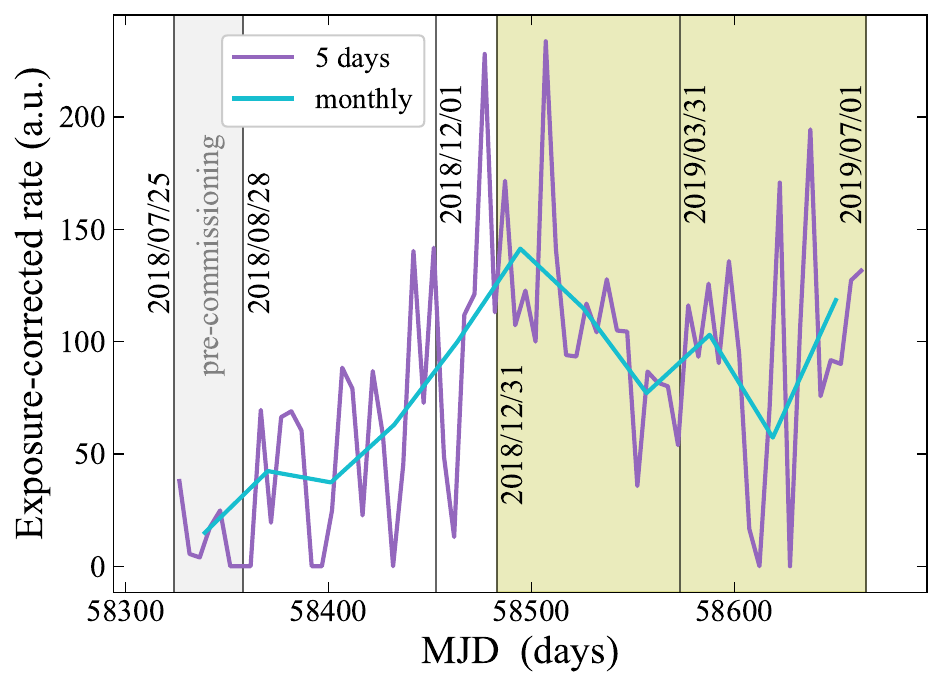}
\caption{Detection rate of bursts during CHIME/FRB Catalog 1 period.  The solid curves represent the exposure-time corrected detection rates averaged over 5 days (purple) and 1 month (light blue) in arbitrary units. The gray-shaded vertical stripe represents the pre-commissioning phase, where the CHIME system configuration was different from the remaining period. The yellow-shaded region represents the hypothetical observation period that we adopt for this work. }
\label{fig:rate}
\end{figure}

\begin{figure*}
\centering
\includegraphics[width=0.87\textwidth]{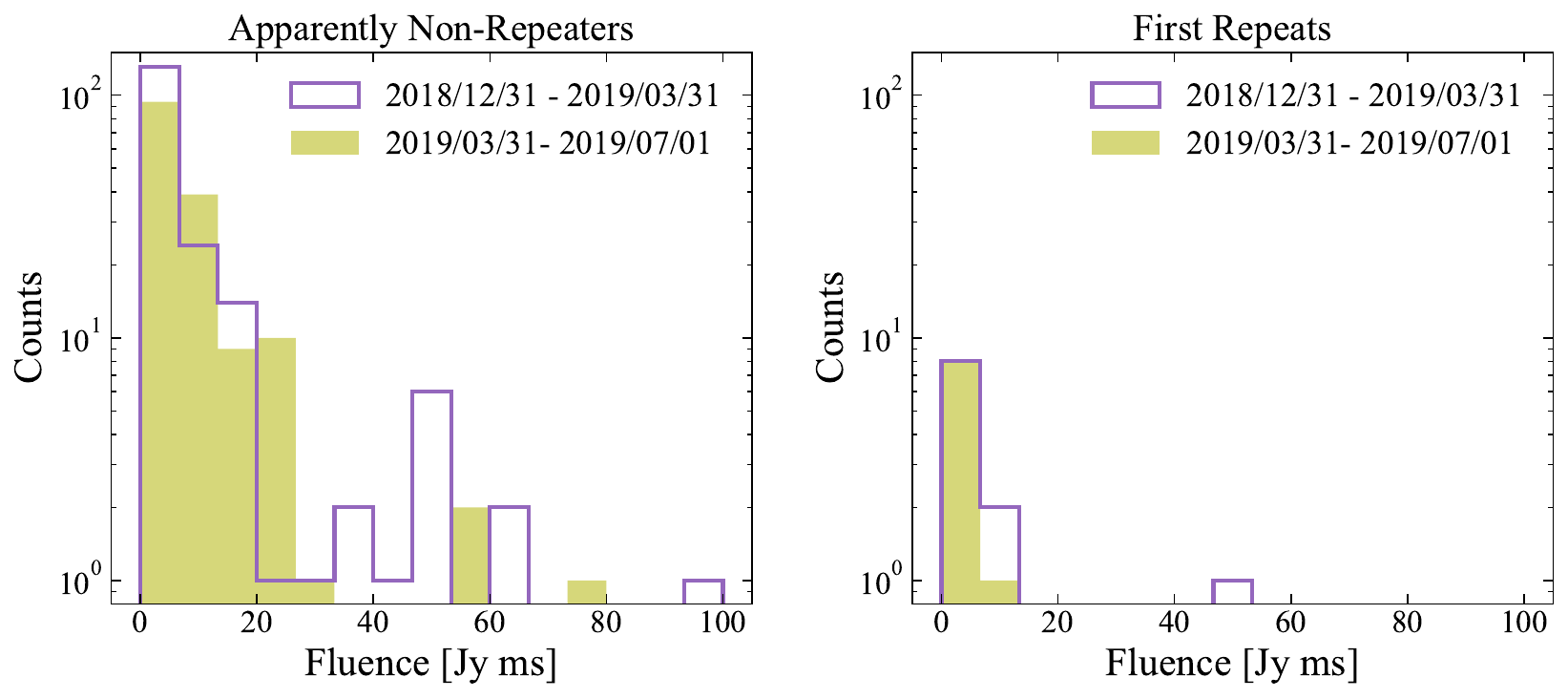}
\caption{Fluence distribution for bursts detected during the latter half of CHIME/FRB Catalog 1 period.  Fluence distributions for apparently non-repeating bursts (left) and first-detected repeater events (right). Each panel compares the fluence distribution of bursts detected from two different time intervals. The first time interval spans from December 31, 2018, to March 1, 2019, while the second time interval covers March 1, 2019, to July 1, 2019.}
\label{fig:fluence}
\end{figure*}

Fig.~\ref{fig:rate} shows the detection rate comparison between the early and late halves of Catalog 1. We found that the detection rate during the early half of the Catalog 1 period was somewhat unstable and lower than the late half. 
To mitigate any impact of varying sensitivity on our results, we chose an observation period that starts on December 31, 2018 and ends on July 1, 2019. However, even during this period, we observed fluctuations in the detection rate.
To further investigate the impact of sensitivity on the fraction of repeaters ($f_{\rm r}$), we examined the fluence distribution of CHIME apparently non-repeating and repeat bursts detected from two different time intervals, as shown in Fig.~\ref{fig:fluence}. We confirm that the fluence distribution remains comparable although the statistics for repeating bursts are low, indicating that the sensitivity variability does not lead to a change in $f_{\rm r}$.
This revised observation period helps us select a more consistent and reliable dataset for our analysis. We also checked that the decreasing trend observed in the apparently non-repeating source detection rate at late observation hours remains consistent, regardless of the starting date chosen for hypothetical CHIME observation.

\section{Exposure Time \& Observed source count}
\label{app:observed sc}

To derive the observed source counts from the cataloged FRBs, we define the fraction of accumulated exposure time to the total operational time spent up to the point of FRB detection, denoted as
$f_{{\rm exp},\,i} \equiv T_{{\rm exp},\,i}/T$, where $T_{{\rm exp},\,i}$ represents the total on-source exposure time accumulated on the $i$-th FRB in the catalog, and $T$ corresponds to the total operational time. CHIME/FRB Catalog 1 provides the total exposure time accumulated on each FRB's position during the entire Catalog 1 period. Since the CHIME exposure time linearly scales with the operational time, we scale the total exposure time during the Catalog 1 period by multiplying it by the fraction of operation time accumulated until the time of detection for each FRB and the total operational period. This allows us to estimate the specific on-source exposure time until the detection of each FRB since $T_0$.

The cumulative source count until a given operational time $T$ is defined as the total sum of events weighted by $w_i = 1/f_{{\rm exp},\, i}$. This essential correction ensures that the counts accurately reflect the effect of exposure time and provide a reliable estimate of the occurrence of apparently non-repeating and repeating FRB events.
Initially, each burst is counted as an apparently non-repeating FRB source until it repeats. Once it does, it is reclassified as a repeating source and subsequently removed from the apparently non-repeating source count. This ensures that each burst is accurately categorized and counted either as an apparently non-repeating or repeating source based on its behavior during the observation period, resulting in a reliable cumulative source count. 
To determine the uncertainties associated with the weighted source counts, we scaled the Poisson error ($\approx\sqrt{N}$ for $N\gg 1$) by the standard deviation of weight $w_i$.

\begin{figure}
\centering
\includegraphics[width=0.45\textwidth]{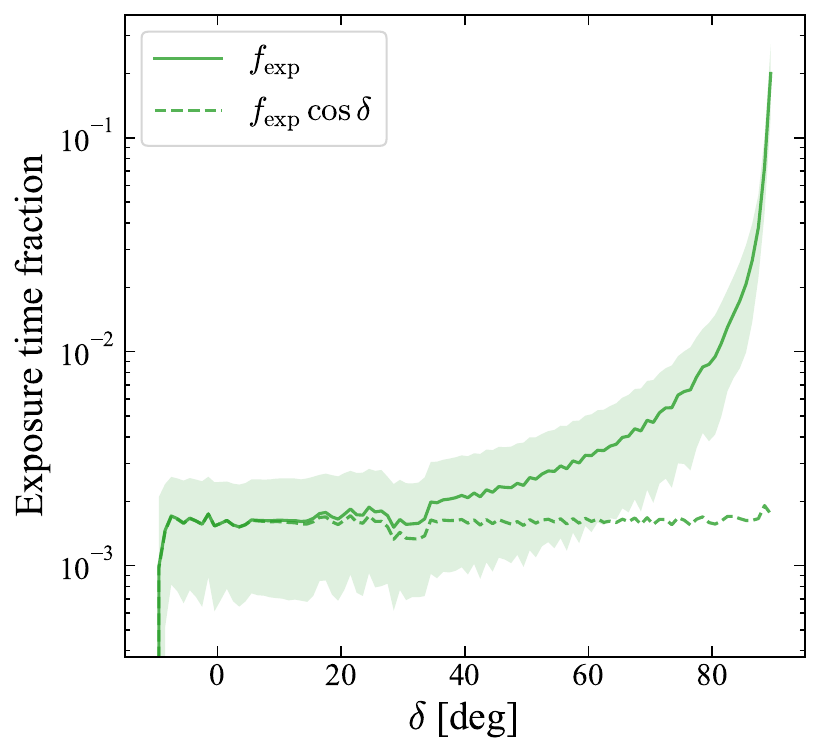}
\caption{Average CHIME exposure time fraction as a function of declination. The solid line shows the exposure time normalized by the total operation period. The dashed line represents the same but after multiplying by $\cos\delta$, which remains constant over $\delta$. This demonstrates that the declination dependence arises purely from the geometric effect of varying sky coverage at different declinations. The decrease in mean exposure at $27^\circ < \delta < 34^\circ$ is a result of a temporary failure of CPU \citep{chime-repeaters23}, which is accounted for in the analysis.}
\label{fig:exposure}
\end{figure}

\section{Monte Carlo simulations}
\label{app:simulation}

We employ a Monte Carlo simulation based on a two-population model that includes both repeating and truly non-repeating sources. For repeating sources, we model the distribution of time intervals between consecutive bursts using the Weibull distribution \citep{opperman18}, which is characterized by two parameters: the mean repeating rate denoted as $r$, and the degree of wait-time clustering represented by $k$. When $k<1$, the distribution exhibits temporally clustered bursts, while in the limit of $k\gg 1$, the burst distribution becomes nearly periodic. The special case with $k=1$ corresponds to the Poisson distribution. To determine the mean repeating rate ($r$), we sample from the differential rate distribution $dN/dr \propto r^{-q}$, where $q$ ($>0$) is the power-law index and $r_{\rm min}$ is the minimum repeating rate. Both $q$ and $r_{\rm min}$ are free parameters in our analysis. We set the maximum repeating rate $r_{\rm max}$ to a fixed value of $10^{0.5}\ {\rm hr^{-1}}$ constrained by the repeating rate of known active repeaters \citep{lu20,ai21}. In this study, we assume that the lifetime of repeaters is much longer than the observational timescale, and thus the evolution of repeaters is effectively negligible.

We generate a large sample of repeating sources ($N_{\rm MC}=10^5$) by the Monte Carlo technique and they are uniformly distributed in the cosine of declination (at $-11^\circ < \delta < 90^\circ$), as most FRBs are extragalactic. For each source, we generate burst time series according to the Weibull distribution. Time intervals between consecutive bursts, $T_{\rm wt}$, are drawn from the inverse cumulative distribution function (CDF) of the Weibull distribution: 
\begin{equation}
    T_{\rm wt} = \frac{[-\ln {\cal U}(0,1)]^{1/k}}{r\,\Gamma(1+1/k)}\ ,\nonumber
\end{equation}
where $r$ is the rate parameter drawn from the rate distribution, $\Gamma$ is the gamma function, and ${\cal U}(0,1)$ is a random number sampled from a uniform distribution between 0 and 1. The first burst time is determined by $T_{0}={\cal U}(0,T_{\rm wt})$, and the $i$-th burst time is given by $T_{i}=T_{i-1}+T_{\rm wt}$ \citep{gardenier19,ai21}.

To simulate the CHIME observation setup, we utilize an exposure map for CHIME/FRB Catalog 1 period integrated from August 28, 2018, to July 1, 2019. For each declination bin, we take the mean of the total exposure time over longitude, $T_{\rm exp}(\delta)$, and then divide it by the total operation period during Catalog 1, $\Delta T$ (which was 308 days). This yields the fraction of exposure time, denoted as $f_{\rm exp}(\delta) = T_{\rm exp}(\delta) / \Delta T$. Fig. \ref{fig:exposure} shows the mean exposure 
time fraction as a function of declination for upper transit. 
At $\delta >70^\circ$, sources pass the detector twice per day. We use the same exposure time function for the lower and upper transits as they are similar \citep{chime-catalog1}.

Each simulated repeating source is then observed by the hypothetical CHIME detection time window of $T_{\rm exp}(\delta)$ once a day if they are at $\delta < 70^\circ$ during the upper transit. If they are in the circumpolar sky locations at $\delta > 70^\circ$, they are observed twice a day in both upper and lower transits. Within the simulated time windows, we count both apparently non-repeating and repeating events, 
correcting for the exposure time fraction as previously described when deriving the observed source count. The modeled source counts are normalized by the number of simulated sources $N_{\rm MC}$, allowing each source count model template to be linearly scaled by an arbitrary total number of repeating sources, $N_{\rm r}^{\rm tot}$. 

Genuine non-repeating sources are characterized by distinct progenitors producing a single FRB during their lifetimes. As they are always within the telescope's field of view with a constant cosmological event rate, we model the cumulative apparently non-repeating source count of truly non-repeating sources as $N_{\rm nn}=\dot{N}_{\rm nn} T$, where $\dot{N}_{\rm nn}$ is the non-repeating burst rate density, a free parameter in our model. The cumulative source counts for repeating sources and apparent non-repeating sources are given by $N_{\rm rr}(T;~r_{\rm min},~k,~q,~N_{\rm r}^{\rm tot})$ and  $N_{\rm rn}(T;~r_{\rm min},~k,~q,~N_{\rm r}^{\rm tot})+N_{\rm nn}(T;~\dot{N}_{\rm nn})$, respectively.

\section{MCMC fitting}
\label{app:mcmc}

Since our model implementation is purely numerical, a formal fit to the data requires interpolation among a grid of pre-calculated model templates. We generate the templates over a wide range of parameter sets ${\bf \theta}=(r_{\rm min},~k,~q)$ using a uniform grid spacing 
for a fixed array of operation time bins $T$, which are assumed to be contiguous. In this study, we take $T$ grids such that the template is well-sampled over the CHIME Catalog 1 operation period of interest up to $T\sim4400$ hr.
For a given set of $\theta$ and $T$, the model is calculated by performing linear interpolation on the $\theta$ and $T$ grid. This enables us to obtain a comprehensive model representation for various parameter combinations and operation times. 

\begin{figure*}
\centering
\includegraphics[width=\textwidth]{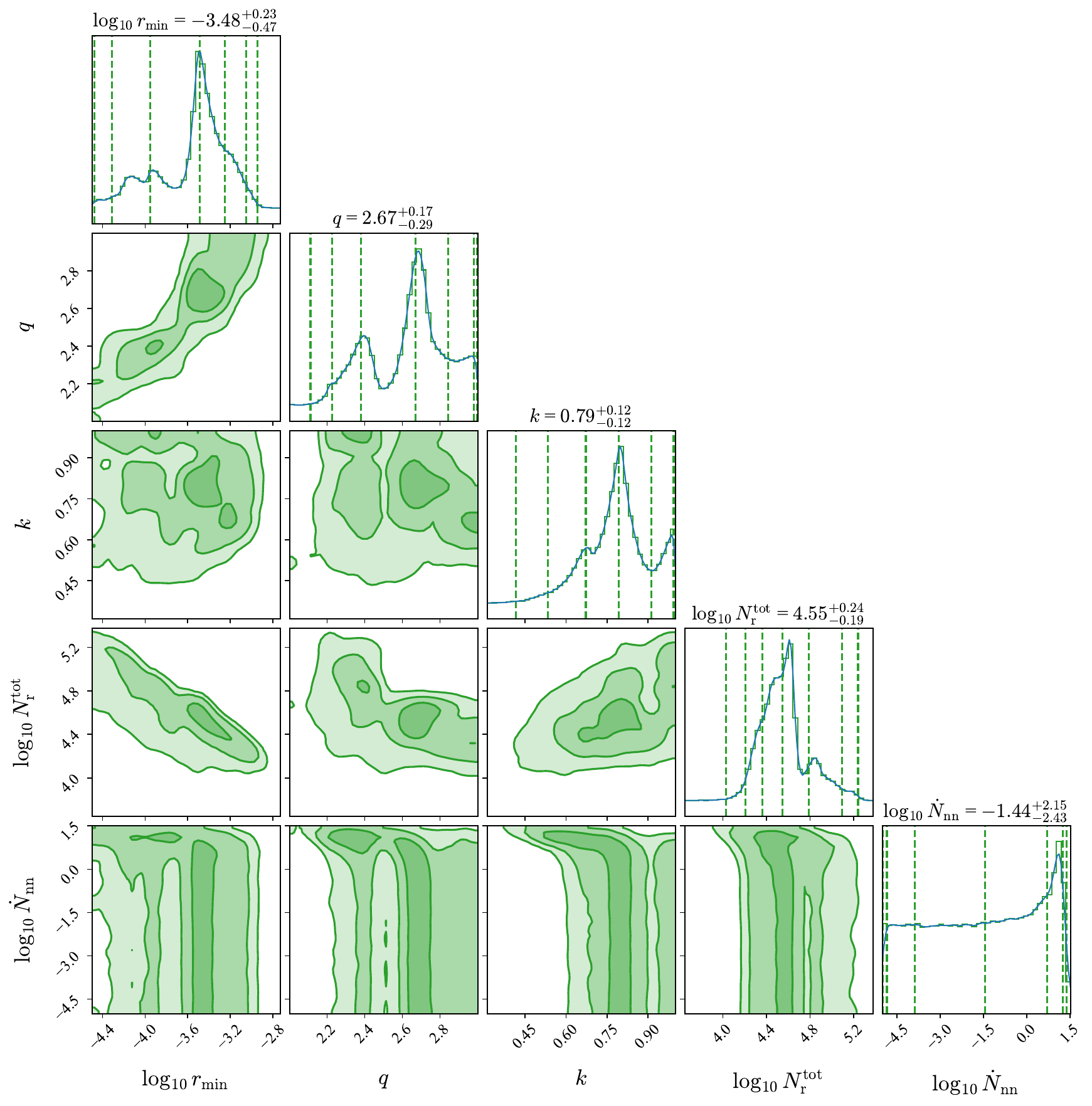}
\caption{MCMC corner plot for the double population model.  The posterior distribution of MCMC samples of parameters $(r_{\rm min},~k,~q,~N_{\rm r}^{\rm tot},~\dot{N}_{\rm nn})$ constrained by CHIME source counts. Vertical dashed lines in marginal distributions indicate CDF values of 0.002, 0.023, 0.159, 0.5, 0.841, 0.977, 0.998 (from left to right). Titles display the median (CDF = 0.5) and 68\% confidence interval (1$\sigma$ error range).}
\label{fig:mcmc_corner}
\end{figure*}

The best-fit parameters are determined using Markov Chain Monte
Carlo (MCMC) sampling with {\fontfamily{qcr}\selectfont emcee}, a
Python-based affine invariant sampler \citep{emcee}.  The
likelihood function is $\ln\left({\cal L}\right)=-\chi^2/2$.  We adopt a flat prior
distribution for all of our parameters in the uniform priors: $\log_{10}(r_{\rm min}/{\rm hr^{-1}})\in[-4.5,-2.5]$, $k\in[0.3,1.0]$, $q\in[2,3]$, $\log_{10}(N_{\rm r}^{\rm tot})\in[3,6]$, and $\log_{10}(\dot{N}_{\rm nn}/{\rm hr^{-1}})\in[-5,2]$. Even after extending the above prior ranges, we have confirmed that the results remain unchanged and robust.
We adopted the data set comprising of $17$ data points at $T/({\rm hr})\in \{20,~40,~80,~160,~320,$ $640,~1000,~1800,~2600,~3400,~4200\}$ for the apparently non-repeating source count and $T/({\rm hr})\in\{640,~1000,~1800,~2600,~3400,~4200\}$ for the repeating source count. Importantly, we confirmed that the different choices of data sets do not significantly affect the MCMC results.

Fig.~\ref{fig:mcmc_corner} displays the posterior distributions of parameters. The slight bimodality observed in the $q$ distribution can be attributed to the coarse grid used in the model templates. The calculated reduced $\chi^2$ value is $\chi^2/$dof$=11.67/12=0.97$, where the degree of freedom (dof) is given by the number of data points ($17$) minus the number of free parameters ($5$).

\section{True number density of repeaters}
\label{app:number density}
The volumetric source density of repeating FRBs is derived by the integration of the energy function over the energy within a specific observational period.
The volumetric rate density of CHIME repeating FRB sources at the faint end of the energy function  \citep{hashimoto22} is $\sim 10^{4}$ Gpc$^{-3}$ yr$^{-1}$ $\Delta\log E^{-1}$.
Therefore, the volumetric source density of CHIME repeaters is $\sim 5.9 \times 10^{3}$ Gpc$^{-3}$ within the operation period of the CHIME Catalog 1, i.e., 0.59 yr  \citep{hashimoto22}. 
We note that this density accounts for only the observationally identified repeaters.
Our MCMC simulations parameterize the total number of repeating FRB sources, $N_{\rm r}^{\rm tot}$ which include sources yet to be detected with CHIME.
The best-fit value in our analysis is $\log N_{\rm r}^{\rm tot}=4.55^{+0.24}_{-0.19}$. 
Therefore, $N_{\rm r}^{\rm tot}$ is $(3.2^{+2.3}_{-1.2}\times 10^{4})/15 \sim (2.1^{+1.5}_{-0.8}) \times 10^{3}$ times larger than the number of identified repeaters in the CHIME Catalog 1. 
The true volumetric number density of repeating FRB sources, including yet-to-be-detected ones, would be $(5.9 \times 10^{3}) \times (2.1^{+1.5}_{-0.8}) \times 10^{3} \sim (1.2^{+0.9}_{-0.4}) \times 10^{7}$ Gpc$^{-3}$. 

The volumetric birth rate of repeating sources is $\sim(1.2^{+0.9}_{-0.4}) \times 10^{3}\,(\tau_{\rm age}/10^4\ {\rm yr})^{-1}$ Gpc$^{-3}$, where $\tau_{\rm age}$ is the lifetime of repeating sources. 
Although the lifetime of repeaters is unknown, to be compatible with the most abundant progenitors, core-collapse supernovae with rate density of $\sim 10^{5}$ Gpc$^{-3}$ yr$^{-1}$ \citep{taylor14}, the lifetime of repeaters need to be $\tau_{\rm age}\gtrsim10^2$ yr.

\section{ASSESSMENT OF MODEL
FITTING}
\label{app:assessment}

In \S \ref{sec:model}, we determined that the minimum repeating rate for repeaters is roughly $10^{-3.5}$ times per hour. Yet, a natural question arises concerning potential biases introduced by the observational timespan, which coincides with this repetition timescale.

\begin{figure}
\centering
\includegraphics[width=0.48\textwidth]{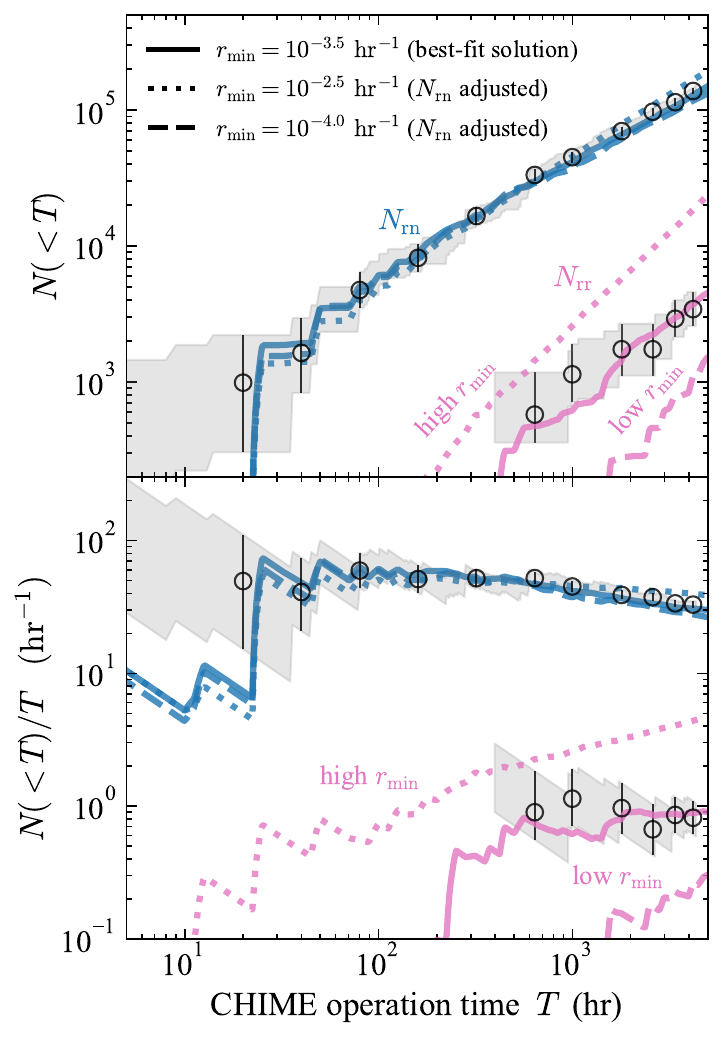}
\caption{Similar to Figure \ref{fig:sc_fit} but additionally including examples where we vary the minimum repeating rate and adjust the non-repeating source counts to align with the observed counts while keeping the other parameters fixed at their best-fit values. The dotted lines depict the scenario with a higher repetition rate, $r_{\rm min}=10^{-2.5}\ {\rm hr^{-1}}$, while the dashed lines represent the model with a lower repetition rate of $r_{\rm min}=10^{-4.0}\ {\rm hr^{-1}}$. }
\label{fig:sc_fit_different_rates}
\end{figure}

Figure \ref{fig:sc_fit_different_rates} illustrates how varying the minimum repeating rate influences the evolution of source counts. The solid lines represent the best-fit model as shown in Figure \ref{fig:sc_fit}, while the dashed and dotted lines depict scenarios where we vary the parameter $r_{\rm min}$ and adjust the modeled non-repeater source count to align with observations (i.e., adjust the normalization $N_{\rm r}^{\rm tot}$) while keeping other parameters the same as their best-fit values. A higher repeating rate (dotted lines) exceeds the observed repeating FRB source count, while a lower repeating rate (dashed lines) falls short (see also \citealt{james23} for a similar discussion on total source counts). This demonstrates the close correlation between the determination of $r_{\rm min}$ and the ratio of the repeater to non-repeater source counts. Hence, we confirm that the inferred minimum repeating rate remains unaffected by the observation period.

\begin{figure}
\centering
\includegraphics[width=0.48\textwidth]{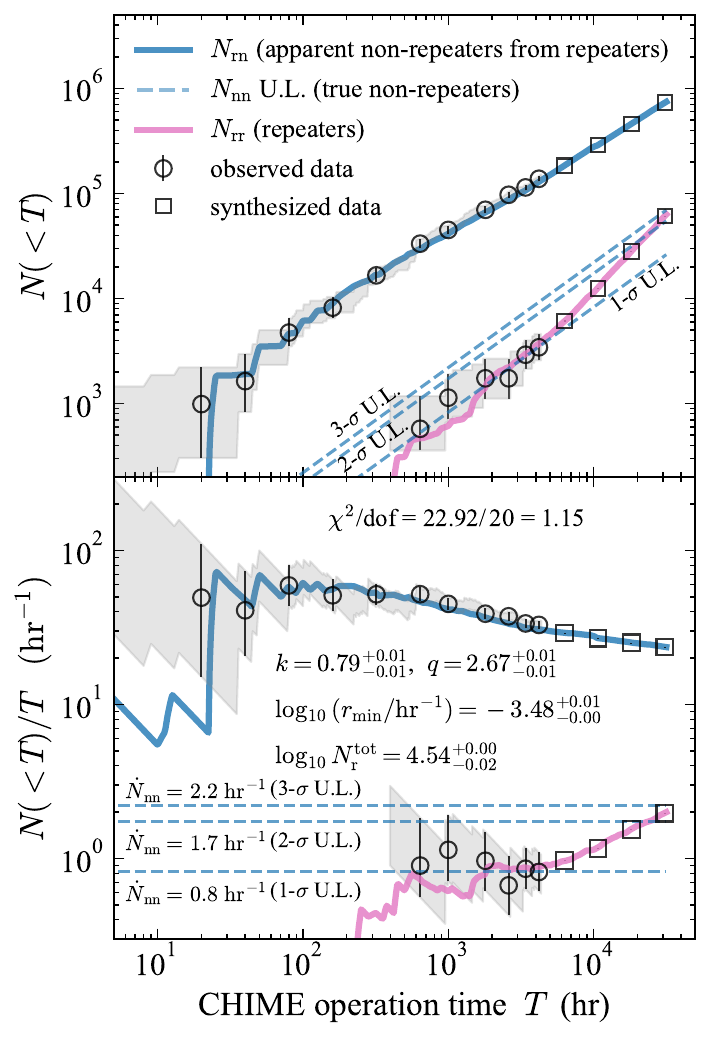}
\caption{Similar to Figure \ref{fig:sc_fit} but for the model fit to the synthesized dataset with an extended operational time. The mock data, simulated at longer operational timescales ranging from $10^{3.5}$ to $10^{4.5}$ hours, is marked by black square markers (see Figure \ref{fig:mcmc_corner_test} for the MCMC posterior distribution of each parameter). }
\label{fig:sc_fit_test}
\end{figure}

\begin{figure*}
\centering
\includegraphics[width=\textwidth]{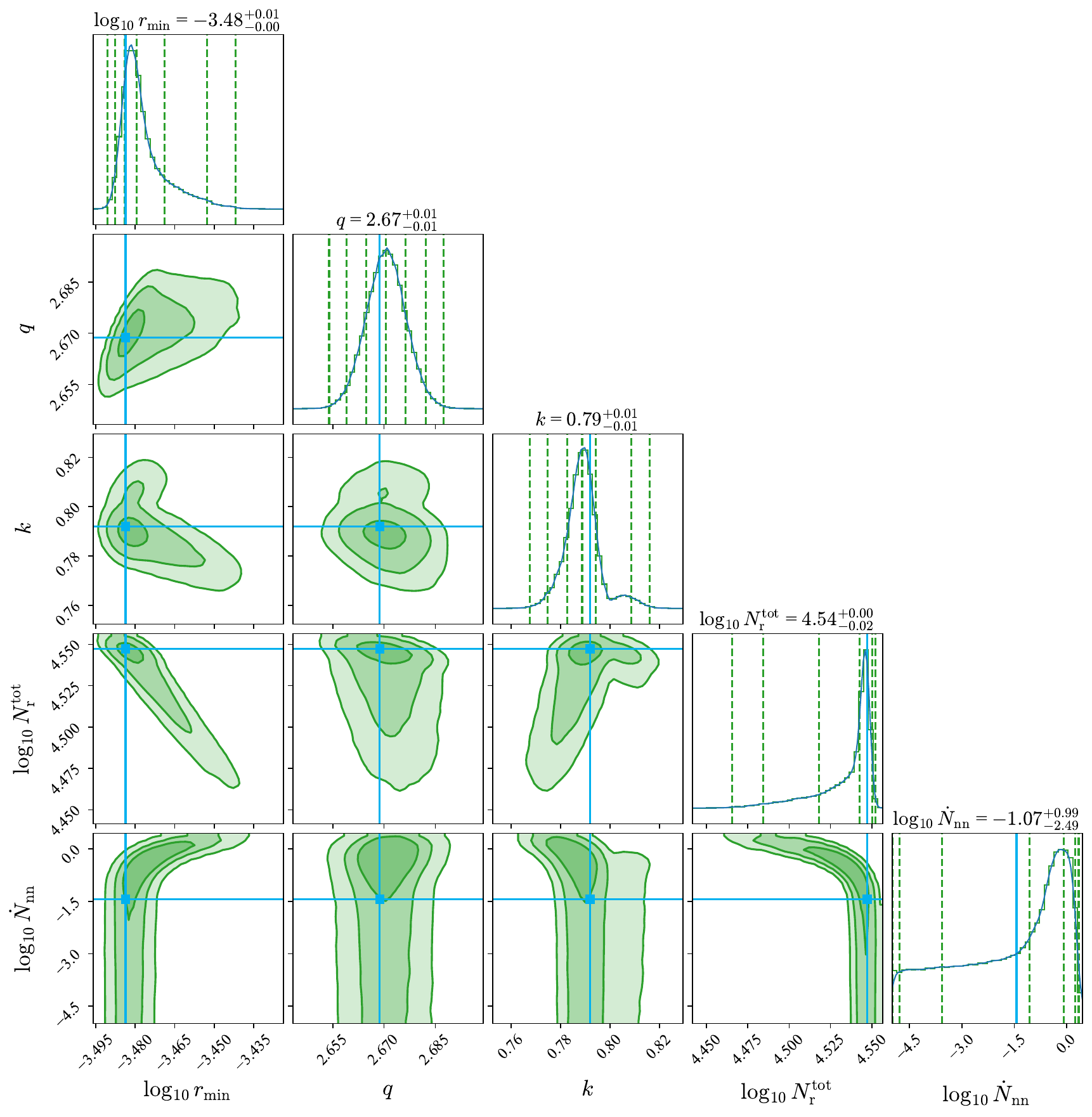}
\caption{MCMC posterior distribution similar to Figure \ref{fig:mcmc_corner}, but for the model fit to the synthesized dataset with an extended operational time (see Figure \ref{fig:sc_fit_test} for the corresponding modeled source counts). The horizontal and vertical light blue lines denote the underlying true parameters, derived from the best-fit results obtained in our main analysis in \S \ref{sec:model} (Figures \ref{fig:sc_fit} and \ref{fig:mcmc_corner}). The posterior distributions consistently align with the assumed parameters within 2-$\sigma$ uncertainties, demonstrating that our model fit accurately recovers the true parameters.}
\label{fig:mcmc_corner_test}
\end{figure*}

Additionally, we generate an extended synthetic dataset with an operational duration of approximately $10^{4.5}$ hours to simulate non-repeating and repeating source counts. Initially, we extrapolate the model using the best-fit parameter set derived from the primary fitting analysis of the observed CHIME dataset, computing model values at extended times. Subsequently, we replace these extended-time model values with new values drawn from a normal distribution centered around the original model values, with a standard deviation determined by Poisson noise. We then assign errors to the replaced dataset using Poisson noise. Figure \ref{fig:sc_fit_test} displays the synthesized dataset at extended times, combined with the observed dataset identical to that used in the main analysis. Finally, we employ MCMC to fit the model to the synthetic data. Figure \ref{fig:mcmc_corner_test} (and see Figure \ref{fig:sc_fit_test} for the model curve with the new best-fit parameters) demonstrates that the marginalized best-fit values effectively recover the ``true'' (initial best-fit) underlying parameters used to generate the dataset, within the 2-$\sigma$ uncertainties. Furthermore, we confirm the robustness of the fitting results across different error realizations, ensuring the accuracy and reliability of our fitting methodology.

Therefore, we confirm that the inferred minimum repeating rate remains robust and unaffected by the observation period. Qualitatively, the minimum repeating rate is shaped by the interplay between the timescale of the diminishing trend in non-repeaters, the emergence of repeating sources, and their source count ratios. Analysis of the current CHIME Catalog 1 data suggests an approximate timescale on the order of $3000$ hours for this interplay.

%%%%%%%%%%%%%%%%%%%%%%%%%%%%%%%%%%%%%%%%%%%%%%%%%%

% Don't change these lines
\bsp	% typesetting comment
\label{lastpage}
\end{document}